&pdflatex
\documentclass[3p]{elsarticle}
\pdfoutput=1

\usepackage{url}
\usepackage{amssymb}
\usepackage{amsmath}
\usepackage{amsthm}
\usepackage{nicefrac}
\usepackage{tabu}
\usepackage{algpseudocode}
\usepackage[section]{algorithm}

\begin{document}

\let\originalleft\left
\let\originalright\right
\renewcommand{\left}{\mathopen{}\mathclose\bgroup\originalleft}
\renewcommand{\right}{\aftergroup\egroup\originalright}

\newcommand{\tablestrutsize}{3.00cm}

\newcommand{\eqnspace}{2ex}

\newcommand{\cpp}{{\nolinebreak C\texttt{++} }}

\newcommand{\BigO}[1]{\mathop{}\!O{\left(#1\right)}}

\newcommand{\Del}[1]{\operatorname{Del}\left(#1\right)}
\newcommand{\Vor}[1]{\operatorname{Vor}\left(#1\right)}
\newcommand{\Conv}[1]{\operatorname{Conv}\left(#1\right)}
\newcommand{\DelS}[1]{\operatorname{Del}|_{\Sigma}\left(#1\right)}
\newcommand{\DelV}[1]{\operatorname{Del}|_{\Omega}\left(#1\right)}
\newcommand{\TS}{\mathcal{T}|_{\Sigma}}
\newcommand{\TV}{\mathcal{T}|_{\Omega}}
\newcommand{\reledge}{h_{r}}
\newcommand{\MAD}{\operatorname{MAD}\left(\theta_{f}\right)}

\newtheorem{proposition}{Proposition}
\newtheorem{corollary}{Corollary}
\newtheorem{lemma}{Lemma}
\newdefinition{definition}{Definition}
\newproof{pf}{Proof}

\begin{frontmatter}

\title{Multi-resolution unstructured grid-generation for geophysical applications on the sphere\tnoteref{tnote1}}
\tnotetext[tnote1]{This paper was presented as a research note at the 24th International Meshing Roundtable, University of Texas at Austin, October, 2015.}

\author[mit,nasa]{Darren Engwirda\corref{cor1}}
\ead{engwirda@mit.edu}


\address[mit]{Department of Earth, Atmospheric and Planetary Sciences, Massachusetts Institute of Technology, 54-918, 77 Massachusetts Avenue, Cambridge, MA 02139-4307, USA}
\address[nasa]{NASA Goddard Institute for Space Studies, 2880 Broadway, New York, NY 10025 USA}

\begin{abstract}
An algorithm for the generation of non-uniform unstructured grids on ellipsoidal geometries is described. This technique is designed to generate high quality triangular and polygonal meshes appropriate for general circulation modelling on the sphere, including applications to atmospheric and ocean simulation, and numerical weather predication. Using a recently developed Frontal-Delaunay-refinement technique, a method for the construction of high-quality unstructured ellipsoidal Delaunay triangulations is introduced. A dual polygonal grid, derived from the associated Voronoi diagram, is also optionally generated as a by-product. Compared to existing techniques, it is shown that the Frontal-Delaunay approach typically produces grids with near-optimal element quality and smooth grading characteristics, while imposing relatively low computational expense. Initial results are presented for a selection of uniform and non-uniform ellipsoidal grids appropriate for large-scale geophysical applications. The use of user-defined mesh-sizing functions to generate smoothly graded, non-uniform grids is discussed.
\end{abstract}

\begin{keyword}
Geophysical Fluid Dynamics \sep Ocean Modelling \sep Atmospheric Modelling \sep Numerical Weather Prediction \sep Voronoi Tessellation \sep Frontal-Delaunay Mesh Generation
\end{keyword}

\cortext[cor1]{Corresponding author. Tel.: +1-212-678-5521.}

\end{frontmatter}



\section{Introduction}
\label{section_introduction}

Planetary climate simulation and numerical weather predication are two of the most computationally challenging tasks currently under consideration. General circulation models typically involve the solution of a large coupled set of non-linear transport equations on very high-resolution spatial grids. The complexity of these models is often significant, with the discrete systems typically involving dozens of independent variables, including fluid momentum, thermodynamic state, such as temperature, salinity and precipitable water, atmospheric aerosols and oceanic biogeochemical tracers. While many such models have previously been developed using structured discretisations of the underlying spherical geometry, recent interest in the use of unstructured general circulation models necessitates the development of effective unstructured mesh generators that cater to this class of numerical scheme. This study investigates the applicability of a recently developed Frontal-Delaunay surface meshing algorithm for this task.

\subsection{Related Work}

\begin{figure}
\centering
\caption{Semi-structured meshing for the sphere, showing a cubed-sphere grid on the left, and an icosahedral grid on the right. Both grids are generated with equivalent mean edge lengths.}

\label{figure_grids}

\bigskip

\includegraphics[width=7cm]{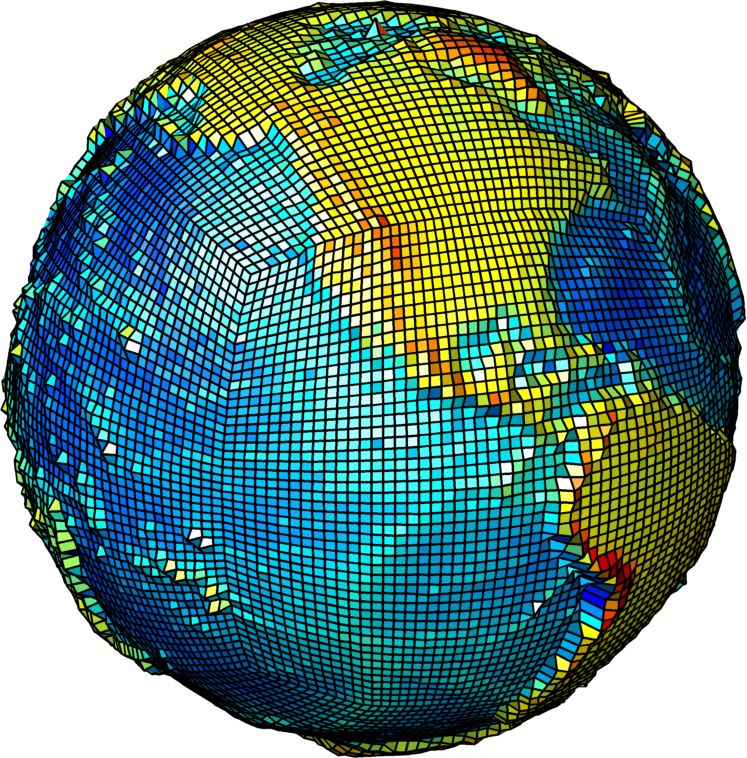}\qquad\qquad
\includegraphics[width=7cm]{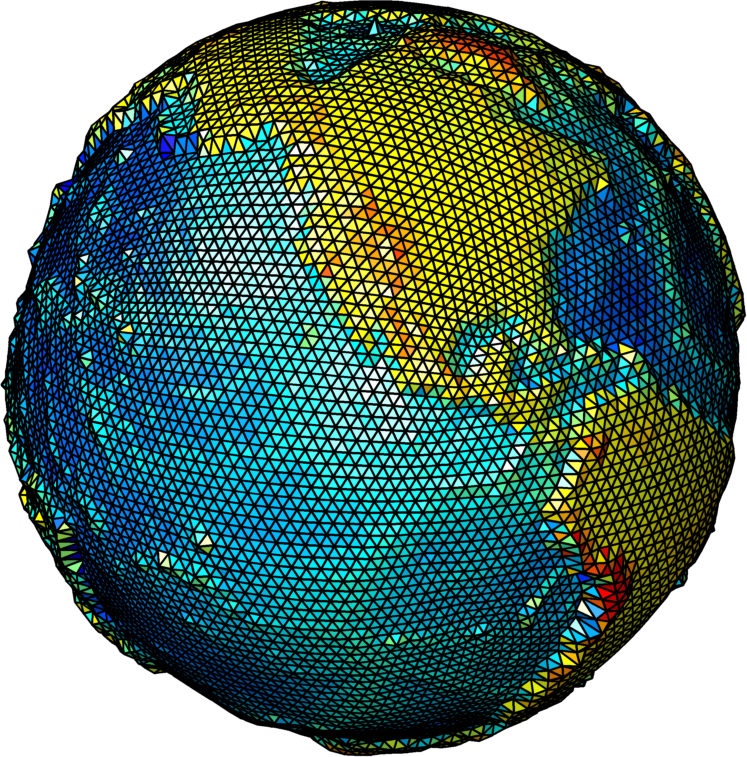}

\end{figure}

While a simple structured mesh for the sphere can be obtained by building a uniform discretisation in spherical coordinates, the resulting `lat-lon' grid is inappropriate for numerical simulation, incorporating a pair of grid singularities at the poles. A majority of current generation general circulation models, (i.e. \cite{adcroft2004implementation,marshall1997finite}), are instead based on a semi-structured discretisation known as the \textit{cubed-sphere}. According to this methodology, the spherical surface is decomposed topologically into an array of six quadrilateral `faces', with each face in turn decomposed into a simple structured curvilinear grid. Each face of the cube is projected onto the adjacent spherical geometry. The cubed-sphere topology is illustrated in Figure~\ref{figure_grids}. In addition to the cubed-sphere configuration, a second class of semi-structured spherical grid can be obtained through icosahedral-type decompositions. In such cases, the primary grid is a triangulation, though the dual polygonal grid is often also used. Icosahedral-type grids are also illustrated in Figure~\ref{figure_grids}. 

Recent general circulations models, specifically the Model for Predication Across Scales (MPAS) \cite{skamarock2012multiscale,ringler2013multi,ringler2008multiresolution}, has focused on the use of locally refined, unstructured polygonal grids to achieve multi-resolution representations of both atmospheric and oceanic dynamics. The current study explores the development of an algorithm for the generation of high resolution spheroidal Delaunay triangulations and associated Voronoi diagrams appropriate for such climate models. Note that in addition to the high-resolution requirements, the sensitivity of the underlying numerical schemes employed in general circulation models necessitates the construction of meshes with near-optimal mesh quality.

\section{A Frontal-Delaunay-Refinement Grid-Generator}
\label{section_algorithm}

The task is to generate very high-quality meshes for planetary atmospheres and/or oceans. Grids are a semi-structured `2.5D' configuration, in which a two-dimensional unstructured grid generated over the spheroidal surface is extruded in the radial direction to form a semi-structured `stack' of fluid layers. The layer-wise grid is set-up to follow the planetary topography -- typically represented as a set of elevations defined from a reference spheroid. In a general form, this reference surface can be expressed in terms of an axis-aligned triaxial ellipsoid:
\begin{equation}
\sum_{i=1}^{3} \frac{(x_{i}-c_{i})^{2}}{r_{i}^{2}} = 1
\end{equation}
\label{equation_ellipsoid}
where the $x_{i}$'s are the Cartesian coordinates in a locally aligned coordinate system, the point $\left(c_{1},c_{2},c_{3}\right)$ is the centre of the ellipsoid, and the scalars $\left(r_{1},r_{2},r_{3}\right)$ are its principal radii.

\subsection{Mesh-size Functions}
Local mesh density can be controlled via a user-specified mesh-size function $\bar{h}(\mathbf{x})$ defined on a background grid. Continuous sizing values are computed using an interpolation scheme. In the current study, mesh-size functions are composed using structured `lat-lon' grids and bilinear interpolation -- providing support for mesh-size functions derived from high-resolution satellite altimetry data \cite{amante2009etopo1}. 

\subsection{Delaunay Mesh Generation}
A high-quality triangular surface mesh is generated on the ellipsoidal reference surface (\ref{equation_ellipsoid}) using a `Frontal-Delaunay' variant of the conventional restricted Delaunay-refinement algorithm \cite{boissonnat03ProvablyGoodSurface,boissonnat05ProvablyGoodMeshing,jamin2013cgalmesh,
cheng2007sampling,Cheng10PiecewiseSmoothMeshing}. This technique is described by the author in detail in \cite{engwirda2015cad,engwirda2014face} and differs from standard Delaunay-refinement approaches in terms of its placement of Steiner vertices. Specifically, the Frontal-Delaunay variant employs various \textit{off-centre} techniques \cite{Rebay93FrontalDelaunay,Ungor09OptSteiner}, designed to position refinement points such that element-quality and mesh-size constraints are satisfied in a locally optimal fashion. Previous studies have shown that such an approach typically leads to substantial improvements in mean element-quality and mesh smoothness. Additionally, it has been demonstrated that this method inherits much of the theoretical robustness of standard Delaunay-refinement techniques -- offering guaranteed convergence, topological correctness, and a minimum angle guarantee.

Given a user-defined mesh-size function $\bar{h}(\mathbf{x})$ and an upper-bound on the element radius-edge ratios $\bar{\rho}$, the Frontal-Delaunay algorithm proceeds to sample the ellipsoidal surface by refining any surface triangle that violates either the mesh-size or element-quality constraints. Refinement is accomplished by inserting a new Steiner vertex at the off-centre point associated with a given element. Refinement continues until all constraints are satisfied. Upon termination, the resulting surface  mesh is guaranteed to contain nicely shaped surface triangles, satisfying both the radius-edge constraints $\rho(f_{i})\leq\bar{\rho}$
and the mesh-size limits $h(f_{i})\leq\bar{h}(\mathbf{x}_{f})$ for all surface triangles $f_{i}$ in the mesh.

As a \textit{restricted} Delaunay-refinement approach, a full three-dimensional Delaunay tetrahedralisation $\Del{X}$ is incrementally maintained throughout the surface meshing phase, where $X\in\mathbb{R}^{3}$ is the set of vertices positioned on the surface of the ellipsoid geometry. The set of restricted surface facets $\DelS{X}$ that conform to the underlying ellipsoidal surface are expressed as a subset of the tetrahedral faces $\DelS{X}\subseteq\Del{X}$. In an effort to minimise the expense associated with maintaining the full-dimensional topological tessellation, an additional \textit{scaffolding} vertex $\mathbf{x}_{s}$ is initially inserted at the centre of the ellipsoid. This has the effect of simplifying the resulting topological structure of the mesh, with the resulting tetrahedral elements forming a simple configuration in which they emanate radially outward from $\mathbf{x}_{s}$. 

\subsection{Mesh Optimisation}
While the Frontal-Delaunay-refinement algorithm typically produces surface triangulations of very high-quality, these tessellations can be further improved through a mesh optimisation step. In this study, a \textit{spring-based} relaxation strategy, adapted from the \texttt{DISTMESH} algorithm of Persson and Strang \cite{Persson04PhD}, is used to optimise vertex positions. Given an initial vertex distribution $\mathbf{x}^{0}$, generated using the Frontal-Delaunay-refinement algorithm described previously, the relaxation procedure seeks updates to the vertex positions in order to achieve an approximate \textit{force-equilibrium} between the mesh vertices
\begin{equation}
\label{equation_springs}
\mathbf{x}^{k+1} = \mathbf{x}^{k} + \Delta t\, \mathbf{F}(\mathbf{x}^{k})\,, \quad
\mathbf{F}_{i}(\mathbf{x}^{k}) = \sum_{j\,\in\,\mathcal{R}_{i}} F_{i,j}(\mathbf{x}^{k})\cdot \hat{\mathbf{e}}_{i,j}\,, \quad
F_{i,j}(\mathbf{x}^{k}) = \beta\,\left(l_{i,j}(\mathbf{x}^{k}) - \bar{h}_{i,j}(\mathbf{x}^{k})\right).
\end{equation}
Here $\mathcal{R}_{i}$ is the set of vertices adjacent to the $i$-th vertex, $l_{i,j}(\mathbf{x}^{k})$ is the length of the edge between the vertices $\mathbf{x}_{i}$ and $\mathbf{x}_{j}$, $\bar{h}_{i,j}(\mathbf{x}^{k})$ is the value of the mesh-size function computed at the associated edge midpoint, and $\hat{\mathbf{e}}_{i,j}$ is the corresponding unit edge direction vector. The spring constant $\beta$ is set to unity in this study. Noting that (\ref{equation_springs}) does not constrain vertices to remain on the ellipsoidal reference surface, an additional projection operator is emloyed within each iteration to enforce the surface constraints exactly. Incremental modifications to the underlying Delaunay mesh topology are enacted throughout the iteration process, ensuring that the mesh remains fully Delaunay.

\subsection{Construction of Voronoi Cells}
Following the generation and optimisation of the surface triangulation, a dual polygonal grid is derived from the associated Voronoi complex. The surface Voronoi diagram contains a grid-cell for each vertex in the underlying surface triangulation, where each grid-cell is a closed, \textit{non-planar} convex polygonal region that connects the centres of the Surface-Delaunay-Balls associated with the triangles adjacent to the central vertex.

\section{Results \& Discussions}
\label{section_results}

\begin{figure}
\centering
\caption{A multi-resolution polygonal grid of the world ocean, generated using the Frontal-Delaunay-refinement algorithm described previously. The non-uniform mesh-size function is computed based on the local gradient in bathymetry, ensuring that regions of rapid change are better resolved.}

\label{figure_ocean}

\bigskip

\includegraphics[width=.975\textwidth]{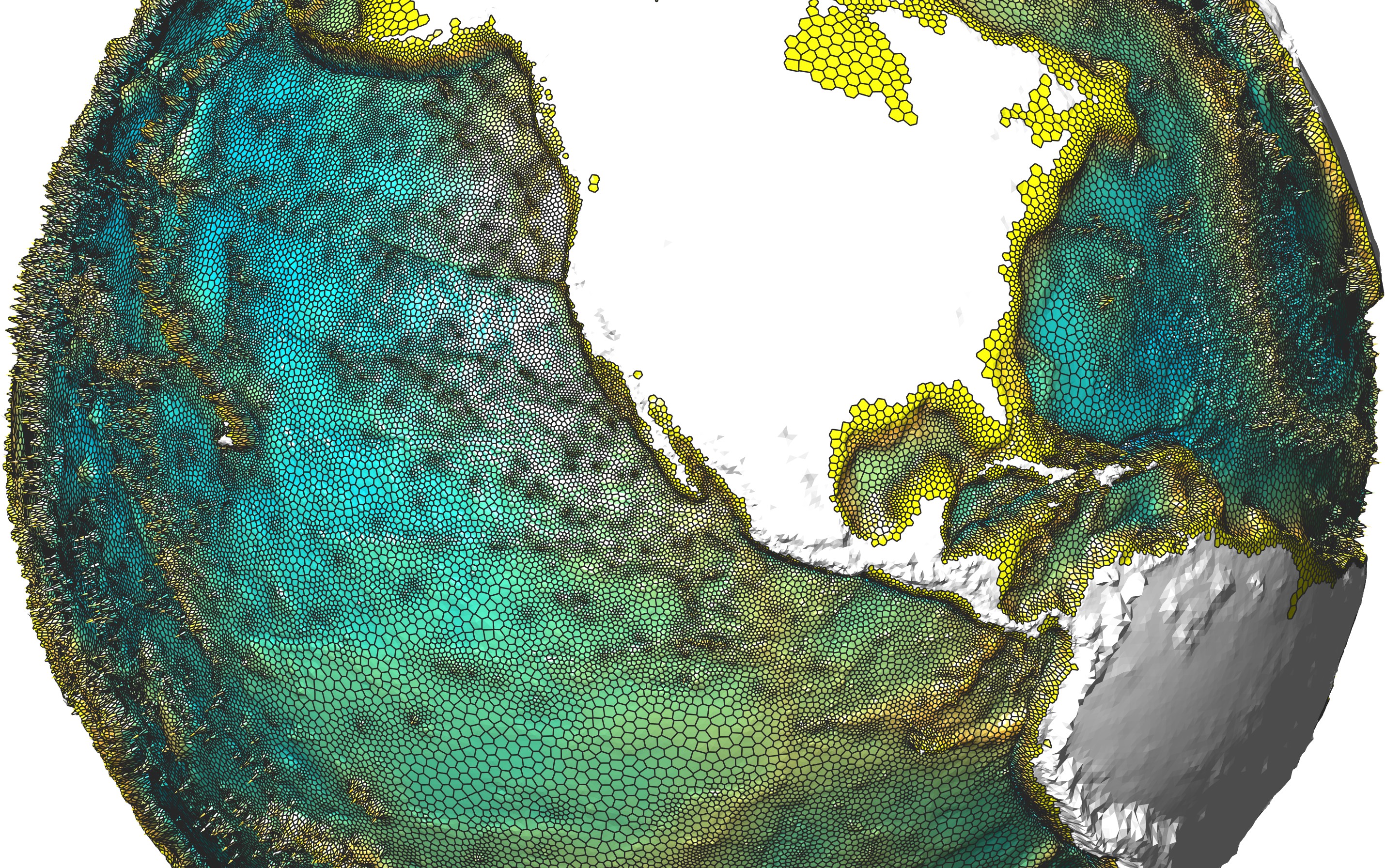}

\end{figure}

Initial investigations have been conducted using the Frontal-Delaunay-refinement approach described previously. A range of very high-resolution spherical grids have been generated using both uniform and non-uniform mesh-size functions. A medium resolution polygonal mesh of the global ocean, generated using a bathymetrically refined mesh-size function is shown in Figure~\ref{figure_ocean}. Future invetigations will analyse the effectiveness of solution-adaptive grid generation schemes, including, for instance, those based on time-averaged distributions of Eddy Kinetic Energy (EKE). 

While the performance of the present algorithm appears to be reasonable, generating in excess of several million elements per minute using standard desktop computing infrastructure, future work will focus on further improvements to performance, including the investigation of dedicated techniques for the computation of Spherical Delaunay Triangulations (SDT's) \cite{renka1997algorithm}, and the development of parallel meshing techniques. A detailed comparison with the existing Parallel Spherical Centroidal Voronoi Tessellation algorithm (MPI--SCVT) of Jacobsen et al.~\cite{gmd-6-1353-2013} will also be conducted.

\bibliography{references}
\bibliographystyle{model1-num-names}

\end{document}